# DIGITAL WATERMARKING TECHNIQUES IN SPATIAL AND FREQUENCY DOMAIN

[1.] Tanmoy SARKAR, [2.] Sugata SANYAL

[1.] Neudesic India Pvt. Limited, Hyderabad, India
[2.] Tata Consultancy Service India Pvt. Limited, Mumbai, India
tanmoy.Sarkar@neudesic.com, sugata.sanyal@tcs.com

Abstract— Digital watermarking [1] is the act of hiding information in multimedia data, for the purposes of content protection or authentication. In ordinary digital watermarking, the secret information is embedded into the multimedia data (cover data) with minimum distortion of the cover data. Due to these watermarking techniques the watermark image is almost negligible visible.
In this paper we will discuss about various techniques of Digital Watermarking techniques in spatial and frequency domains

## INTRODUCTION

With the recent technology advanced people are sharing more and more information among each other's. Some organizations like medicine, military are sharing data with are highly secretive and important. For secure communication people are using cryptography using secret key so that only authenticate receiver can decrypt the message and authentication of message remains intact. But cryptography raised suspicion among attackers and tries to attack the message to get the secretive messages. So, an approach of digital watermarking is used where authenticate multimedia data is embedded into original message. The receiver is then extracts the watermarked image and authenticates its novelty.

## DIGITAL WATERMARKING

Digital watermarking [18] [19] system consists of watermarking encoder and decoder. In watermarking encoder, the digital multimedia data (audio, video, and image), watermarked key and original message put as an input to generated watermarked data.

Types of Digital Watermarking:

1. <u>Visible</u>: This types of watermarking are perceptual to human eyes and can be used for authentication instantly.
2. <u>Invisible</u>: This type of watermarking are not perceptual to human eyes and requires watermarking extraction algorithm.

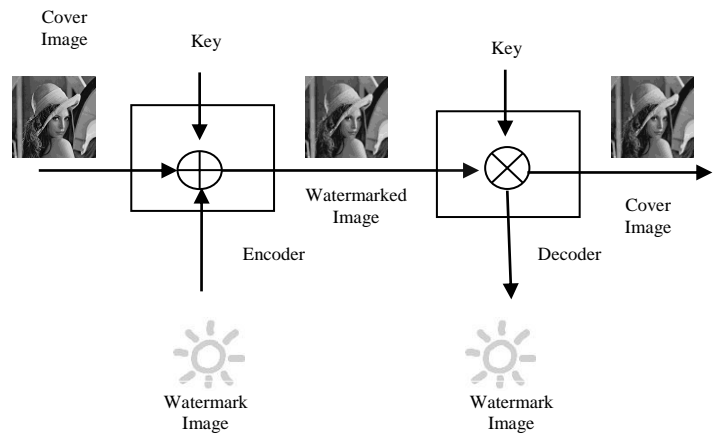

*Figure 1 .Watermarking Technique*

The digital watermarking system can be blind or informed. Blind watermark techniques are independent of cover image. In this technique while transmitting watermarked cover image if any noise is introduced into it then the decoder at the receiving end extract the distorted watermarked image because original cover image is not known to it. In case of informed watermarking techniques the watermarked image is dependent on cover image. While transmitting the hash value of cover image is calculated and incorporate into watermarked image. So that when the image is received at the receiver end the hash value of cover data is calculated and authenticate the data provided the original cover data is with the receiver.

In Digital watermarking systems there are three mutually exclusive parameters to evaluate performance.

1. <u>Quality</u>: Minimum distortion of original image after secret message has been embedded.
2. <u>Capacity</u>: Maximum size of watermark image embedded on cover data.

3. <u>Robustness</u>: Watermarked image should be withstanding any modification attacks.

These three parameters are tightly bound to each other. Trying to improve the performance of one parameter will affect the performances of other two.

This paper deals with invisible watermarking schemes. Here watermarking are not perceptual to human eyes.

Few important properties of Digital Water markings schemes are:

1. <u>Robustness</u>: This property states that the watermark image should resist any possible attach and remain detectable.
2. <u>Fidelity</u>: High fidelity means that the amount of distortion caused by the watermarked image to cover image remains imperceptible to human eyes.
3. <u>Capacity</u>: The number of bits of watermark image can be embedded into cover image without causing much distortion.
4. <u>False Positive Rate</u>: This property state that the identification of watermark image into cover image which doesn't contain actually. Minimum false positive rate helps to identify watermarked cover image easily while decoding.

Applications of Digital Water markings are:

1. <u>Copyright</u>: Using Digital watermarking for copyright purpose helps to protect rights in content distribution. It is used to protect the rights of the owner.
2. <u>Authentication</u>: In order to authenticate the data and to detect tampering of message while transmission digital watermarking is used.
3. <u>Time Stamping</u>: Use of watermarking is this case helps to keep track about when the content was created, last used or last modified.

## STEGANOGRAPHY

Similar to digital watermarking, for embedding secret messages, steganography is used to hide messages in cover data. The basic difference between Steganography and digital watermarking is that in digital watermarking the covert data is related to cover data but in steganography the covert data is not related to cover data.

Steganography is mainly distributed among two approaches: reversible and irreversible [2]. Using reversible technique the receiver can extract both the secret message as well as original cover image but while using irreversible technique the receiver can only extract the secret message from stego image leaving original cover image distorted.

Few irreversible techniques are:

1. Battisti et al [3] approach of data hiding using Fibonacci p-sequence number to reduce stego image distortion than traditional LSB technique.
2. Dey et al [4] [5] [6] proposed an improvement over Fibonacci p-sequence LSB data technique of Battisti et al [1] by decomposing pixel value using two approaches: Prime decomposition and Natural number decomposition technique.
3. Nosrati et al. [7] introduced a method that embeds the secret message using linked-list in RGB 24 bit color image

Some reversible data hiding techniques are:

1. Ni et al. [8] proposes a novel approach of data hiding using histogram shifting of original image
2. Kuo et al. [9] presented a reversible technique that is based on the block division to conceal the data in the image.
3. Tian [10] proposes a reversible data hiding technique using difference expansion.

Similar to cryptanalysis, steganalysis is a technique used to detect steganographic images as mentioned in paper [20].

## DOMAINS USED IN DIGITAL WATERMARKING TECHNIQUES

Spatial Domain Techniques are techniques that operated directly on single pixel of an image.

$$f_i \xrightarrow{T_p(.)} g_i \quad (1)$$

where $f_i$ is the original image, $g_i$ is the modified image and $T_p(.)$ is the spatial operator defined in a neighborhood p of a given pixel.

Frequency Domain Techniques are operated on frequency of an image.

$$f_i \xrightarrow{f_p} I_i \xrightarrow{-f_p} g_i \quad (2)$$

where $f_i$ is the original image, $I_i$ is the modified image after applying frequency transformation $f_p$, $g_i$ is the final modified image after implementing inverse transformation $-f_p$.

## DIGITAL WATERMARKING TECHNIQUES

### Spatial Domain

The simplest method of digital watermarking in spatial domain is using LSB (Least Significant Bit) insertion [17].

**Input:** Cover Image $M_c$ and Watermark Image $M_w$

**Process:**

1. Let $M_{pixel}[i,j]$ is the pixel of Cover Image $M_c$ at position width i and height j.

2. Let Mw is the secret image for watermarking. The length of $M_w$ should be less than cover image $M_c$
3. Loop though the cover image:
$$M_c = M_{pixel}[i,j] - M_{pixel}[i,j] \% 2;$$
$$M_c = M_c + M_w \% 2;$$
$$M_w = M_w/2;$$
4. If the length and width of watermarked image is reached then end.

**Output:** Modified Image Mo having watermark image embedded in it.

The advantage of LSB watermarking technique is its simplicity and the difference is not visible to naked eyes. But this technique has also having lot of disadvantages like LSB encoding is extremely sensitive to any kind of filtering or manipulation. An attack on the watermarked image is very likely to destroy the watermark image. Since this technique is not robust any attack will damage its authentication. From Fig. 2 we can see that after embedding secret message into the cover image there is significant change in original image histogram pattern suggesting it is being distorted.

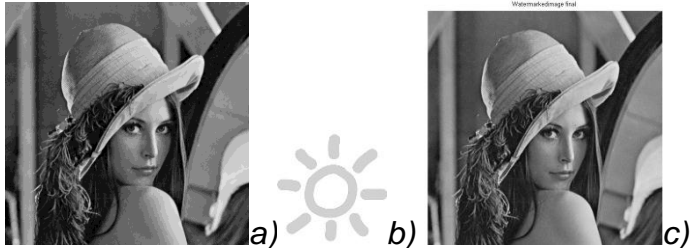

Figure 2. LSB Watermarking Technique (a) Cover Image (b) Image to be embedded (c) Watermarked Image

Tian [10] [11] proposes a digital watermarking technique using difference expansion using LSB. In this method the mean and average value of two neighboring pixel, with small difference value, is calculated. The calculated value is then check to see whether it is satisfying the expandable difference condition eq. (i) and once the condition is passed the new expanded difference is calculated eq. (ii). Finally the watermark image is embedded based on the calculated values. This technique also use location map to store the values to know which difference value have been selected which are used to extract the image at the receiving end. This technique significantly improves the capacity of payload message and visual quality of embedded image.

$$|2 \times h + b| \leq \min(2(255 - l), 2l + 1)) \quad \text{...... (i)}$$
$$h' = 2 \times h + b \quad \text{.......................... (ii)}$$

The algorithm's steps are:
1. Take two adjacent pixel values of x and y
2. Find difference and average values of pixels.
3. Then expand into its binary form and add watermark bit right after most significant bit.

In the fig. 3 we have embedded image (b) into image (a) by using difference expansion. From the histogram of stego image (c) we can see that the watermark image is embedded on the difference of near pixel value which are expandable but the pixel having minimum intensity or zero value are not used much in this process.

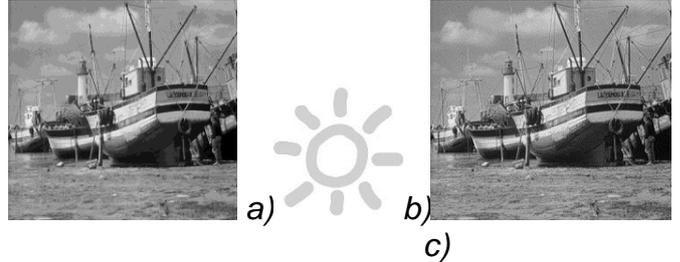

Figure 3. Difference Expansion using LSB (a) Cover Image (b) Watermark Image (c) Watermarked Image

**Frequency Domain**
**Discrete Wavelet Transform**

A lot of research papers used DWT for Digital watermarking techniques [12] [13].The frequency domain transform used here is Haar-DWT. A 2-dimensional Haar-DWT consists of horizontal and vertical operations. Detailed procedures of a 2-D Haar-DWT are described as follows:

Step 1: Pixels are scanned from left to right horizontally, add the value with neighboring pixels and store the sum on the left and the difference on the right as illustrated in Figure 1.4. The operation is repeated until all the rows are processed. Pixel sums represent the low frequency part (denoted as symbol L) while the pixel differences represent the high frequency part 9(denoted as symbol H) of the original image.

Step 2: Pixels are scanned from left to right vertically, add the value with neighboring pixels and store the sum on the left and the difference on the right as illustrated in Figure 4. Repeat this procedure until all the columns are processed. At the end we LL, HL, LH, and HH bands are created. The LL sub-band is the low frequency portion and looks very similar to original cover image.

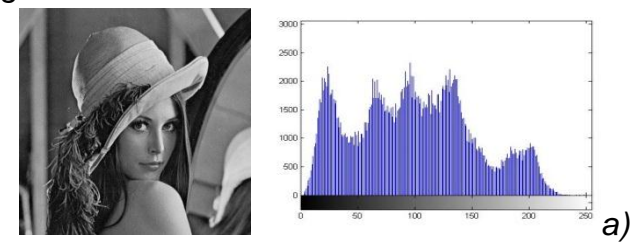

a)

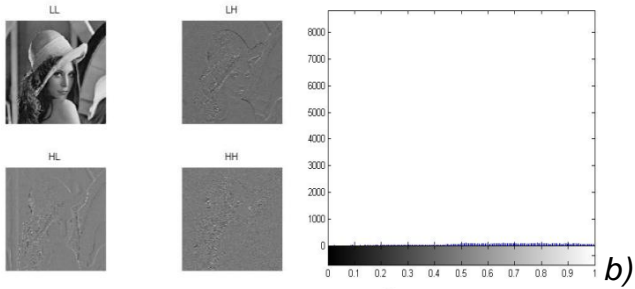
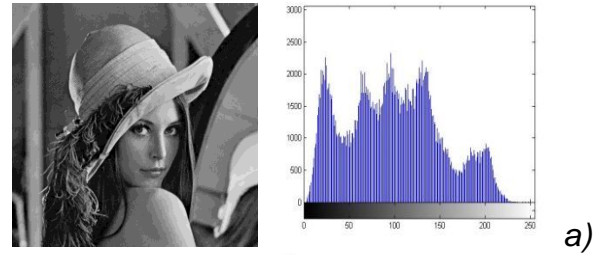

Figure 4. Discrete Wavelet Transformation (a) Cover Image (b) DWT Image (c) Watermark Image (d) Watermarked Image

**Discrete Cosine Transformation**

DCTs [14] [15] [16] are used to convert image processing data from spatial domain to frequency domain into summation of series of cosine waves oscillating at different frequencies. For Image processing 2-D DCT technique is used and is given by:

$$DCT(i,j) = \propto (i) \propto (j) \sum_{x=0}^{N-1} \sum_{y=0}^{N-1} f(x,y) \cos\left[\frac{\pi(2x+1)i}{2N}\right] \cos\left[\frac{\pi(2y+1)j}{2N}\right] \quad (3)$$

where i, j = 1,2,3……,N-1 and α(j) and α(j) is given by :

$$\propto (i) = \propto (j) = \begin{cases} \sqrt{\frac{1}{N}} & For\ i,j=0 \\ \sqrt{\frac{2}{N}} & For\ i,j \neq 0 \end{cases} \quad (4)$$

And inverse transform is given by:

$$f(i,j) = \sum_{x=0}^{N-1} \sum_{y=0}^{N-1} f(x,y) \cos\left[\frac{\pi(2x+1)i}{2N}\right] \cos\left[\frac{\pi(2y+1)j}{2N}\right] \quad (5)$$

Algorithm of DCT:
1. Read the original Input Image
2. Resize the watermark image.
3. Resize the original image and watermark image for efficient partition in blocks.
4. Retrieve the 8×8 sub-blocks of original image and apply DCT to each of them.
5. Apply the watermark into each of these sub-blocks and apply inverse transform.

Figure 5. Discrete Cosine Transformation (a) Cover Image (b) Watermark Image (c) Watermarked Image

**CONCLUSION**

In this paper we described algorithms that belongs to spatial and frequency domain in the digital watermarking techniques. All these techniques are designed to exploit some aspects of the human visual system and made watermark image imperceptible. Many of these techniques rely either on transparency (low-amplitude) or frequency sensitivity to ensure the mark's invisibility. Digital watermarking explores are keep on exploring new methods in these areas and this paper helps to understand and gain knowledge for further researches.

**REFERENCES**
[1.] Ruchira Naskar and R. S. Chakraborty, "Reversible Digital Watermarking: Theory and Practices", Morgan Claypool, USA, ISBN: 978-1627053150
[2.] Tanmoy Sarkar, Sugata Sanyal, "Reversible and Irreversible Data Hiding Techniques" in arxiv.org, arXiv: 1045.2684, 2014
[3.] F. Battisti, M. Carli, A. Neri, K. Egiaziarian, "A Generalized Fibonacci LSB Data Hiding Technique", IEEE 3rd International Conference on Computers and Devices for Communication (CODEC-06) TEA, Institute of Radio Physics and Electronics, University of Calcutta, December 18-20, 2006.
[4.] Sandipan Dey, Ajith Abraham, Sugata Sanyal, "An LSB Data Hiding Technique Using Natural Numbers", Intelligent Information Hiding and Multimedia Signal Processing, IIHMSP 2007. Third International Conference, Kaohsiung, Vol.2, 2007, pp. 473-476.

[5.] Sandipan Dey, Ajith Abraham, Bijoy Bandyopadhyay, Sugata Sanyal, "Data Hiding Techniques Using Prime and Natural Numbers.", Journal of Digital Information Management, vol. 6, no. 6, pp. 463-485, 2008.

[6.] Sandipan Dey, Ajith Abraham, Sugata Sanyal , "An LSB Data Hiding Technique Using Prime Numbers", The Third International Symposium on information Assurance and Security, Manchester, UK, IEEE CS press, pp. 101-108, 2007

[7.] M. Nosrati, R. Karimi, H. Nosrati, and A. Nosrati, "Embedding stego-text in cover images using linked list concepts and LSB technique", Journal of American Science, Vol. 7, No. 6, 2011, pp. 97-100.

[8.] Z. Ni, Y. Q. Shi, N. Ansari and W. Su, "Reversible data hiding," IEEE Transactions on Circuits and Systems for Video Technology, Vol.16, No.3, pp. 354-362, 2006

[9.] Wen-Chung Kuo, Dong-Jin Jiang, Yu-Chih Huang, "A Reversible Data Hiding Scheme Based on Block Division", Congress on Image and Signal Processing, Vol. 1, 27-30 May 2008, pp. 365-369

[10.] Jun Tian, "Reversible Data Embedding Using Difference Expansion", IEEE Transactions on Circuits and Systems for video technology, Vol.13, No. 8, August 2003, pp. 890-896.

[11.] J. Tian, "Reversible watermarking by difference expansion," in Proceedings of Workshop on Multimedia and Security, pp. 19–22, Dec. 2002. 25, 28

[12.] Munesh Chandra, Shikha Pandey "A DWT Domain Visible Watermarking Techniques for Digital Images", pp. 421-427, IEEE 2010.

[13.] Lijing Zhang, Aihua Li "Robust watermarking scheme based on singular value of decomposition in DWT domain", pp. 19-22, Asia-Pacific Conference on Information Processing IEEE 2009.

[14.] A. Bors, I. Pitas "Image watermarking using DCT domain constraints." IEEE International Conference on Image Processing, Lausanne, Switzerland, Sept. 1996, pp. 231-234

[15.] Mei Jiansheng1, Li Sukang1 and Tan Xiaomei "A Digital Watermarking Algorithm Based On DCT and DWT", International Symposium on Web Information Systems and Applications (WISA'09) 2009, pp. 104-107.

[16.] Mohamed Al Baloshi, Mohammed E. Al-Mualla: A DCT-Based Watermarking Technique for Image Authentication. AICCSA, 13-16 May, Amman, Jordan, 2007, pp. 754-760

[17.] Yeuan-Kuen Lee, Graeme Bell, Shih-Yu Huang, Ran-Zan Wang, and Shyong-Jian Shyu, "An Advanced Least-Significant-Bit Embedding Scheme for Steganographic Encoding" PSIVT '09 Proceedings of the 3$^{rd}$ Pacific Rim Symposium on Advances in Image and Video Technology, Springer-Verlag, Berlin, Heidelberg, 2009, pp. 349-360.

[18.] Munesh Chandra, Shikha Pandel, Rama Chaudhary, "Digital watermarking technique for protecting digital images" Third IEEE International Conference on Computer and Information Science and Technology (ICCSIT 2010), pp.226-233.

[19.] A.Z.Tirkel, R.G.van Schyndel, C.F.Osborne. "A Two Dimensional Digital Watermark", DICTA'95, University of Queensland, Brisbane, December 6-8, 1995, pp. 378-383

[20] Tanmoy Sarkar, Sugata Sanyal "Steganalysis: Detecting LSB Steganographic Techniques", IJESM, Vol.4, Issue 2, Page 34